\documentstyle[prl,aps,amssymb,graphicx]{revtex}
\newcommand{\beq}{\begin{equation}}
\newcommand{\eneq}{\end{equation}}

\input{epsf}

\begin{document}

\tolerance 10000

%\twocolumn[\hsize\textwidth\columnwidth\hsize\csname %
%@twocolumnfalse\endcsname

\draft
\title{Interaction among particles with fractionalized quantum numbers in
one dimensional samples}
\author{Bogdan A. Bernevig$^{*,\dagger}$, Domenico Giuliano$^{+,\%}$,
Robert B. Laughlin$^*$}

\address{$^*$  Department of Physics, Stanford University, Stanford,
CA, 94305, USA\\
$^\dagger$ Department of Physics, Massachussets Institute
of Technology, Cambridge, MA, 02139, USA \\
$^+$ Dipartimento di Scienze Fisiche, Universit\`a di Napoli
         ``Federico II'', Monte S.Angelo - via Cintia, I-80126 Napoli, Italy\\
$^\%$ {\bf Present Address} Dipartimento di Fisica, Universit\`a della
Calabria, Arcavacata di Rende (CS), Italy.}

%\twocolumn[
\date{\today}
\maketitle
\widetext

\begin{abstract}
%\vspace*{-1.0truecm}
\begin{center}

The elementary excitations of strongly correlated one-dimensional
electronic systems - spinons and holons - are discussed in an exact
solution of the Haldane-Shastry and Kuramoto-Yokoyama model. We derive
and exactly solve the equation of motion both for the two-spinon
wavefunction and for the one-spinon, one-holon wavefunction. By
solving the equations of motion we find the spinon-spinon and
spinon-holon interaction to be a short-range attraction. The physical
consequences of such an attraction on the spin-form factor and on the
hole-spectral function are also worked out.
\end{center}

\end{abstract}

\pacs{
\hspace{1.9cm}
} 
]

%\narrowtext

\section{Introduction}

Phenomenological Landau's Fermi liquid theory \cite{landau} describes
many of the condensed matter systems. The basic assumption of Landau's
theory is that the spectrum of the interacting Fermi liquid may be
adiabatically deformed to the spectrum of a noninteracting Fermi
gas. Equivalently, one may say that the interaction strength can be
continuously increased from 0 (Fermi gas) to its Fermi Liquid value
without hitting a singular point. The excitations of a Fermi liquid
can be continuously deformed to electrons and holes and are referred to
as quasiparticles and quasiholes, respectively.

In a ``non-Fermi liquid'' such a scenario does not apply. Typical
examples of non-Fermi liquids are the one-dimensional strongly
correlated electronic systems. The elementary excitations of these
systems are collective modes carrying charge 1, but no spin
(``holons''), and modes carrying spin-1/2, but no charge
(``spinons''). Originally found as elementary excitations of spin-1/2
one-dimensional antiferromagnetic spin chains \cite{cloizeaux,fadeev},
spinons appear as localized spin-1/2 spin defects, embedded in an
otherwise featureless spin-singlet spin-liquid state. Spinons and
holons carry ``fractions'' of the quantum numbers of spin waves and/or holes.

In these lectures we analyze the dynamics of spinons and holons in
one-dimensional electron-systems close to half filling. In
particular, we study their interaction in the
framework of simple, exactly solvable model, using a formalism
different from the usual Bethe-ansatz approach \cite{bethe}. Such an
approach treats collective modes of strongly correlated systems
(spinons, holons, etc.) as quantum mechanical particles. In
particular, it is possible to associate a two-body wavefunction to a
two-spinon or to a spinon-holon pair. The corresponding equation of
motion can be straightforwardly solved. The form of the interaction
potential in the ``Schr\"odinger'' operator and the behavior of the
solution as a function of the separation among the particles indicates
a short-range attraction.

In the thermodynamic limit, the spinon-spinon and spinon-holon
attraction drives the low-energy physics of one-dimensional
antiferromagnets and strongly-correlated insulators. It is responsible
for the sharp features at threshold in the spin structure factor, and
in the hole spectral function. These features have been experimentally
observed \cite{alan,zxshen} and they provide an important evidence for
a short-range attractive force.

The notes are organized as follows:

\begin{itemize}

\item In Section II we introduce the Haldane-Shastry model of a
one-dimensional
antiferromagnet. We study in some details its basic features, its ground
state and its elementary excitations;

\item In Section III we discuss the two-spinon wavefunction, together
with the corresponding equation of motion, its solution, and physical
interpretation. We prove the existence of a short-range attraction
among spinons;

\item In Section IV we consider one-holon and one-spinon states,
within the framework of the Kuramoto-Yokoyama model which is the
supersymmetric extension of the Haldane-Shastry model;

\item In Section V we study the one-spinon one-holon wavefunction. We
solve the corresponding equation of motion, and study in detail the basic
features of the solution. Finally, we prove the existence of a short-range
spinon-holon attraction;

\item In Section VI we spell out the physical consequences of the
attraction
between spinons and holons. We trderive its effects on the 
 dynamical spin structure factor in one dimensional
antiferromagnets and on the angle-resolved hole spectral function
in one-dimensional insulators;

\item In Section VII we provide our conclusions.

\end{itemize}

\section{The Haldane-Shastry model of a one-dimensional antiferromagnet
and its elementary excitations}

Spinons appear as elementary excitations of one-dimensional half-odd
spin antiferromagnets.  We explore the dynamics of spinons in the
framework of the Haldane-Shastry (HS) model of a spin-1/2
antiferromagnetic chain \cite{haldane,shastry}.

\begin{figure}
\includegraphics*[width=0.8\linewidth]{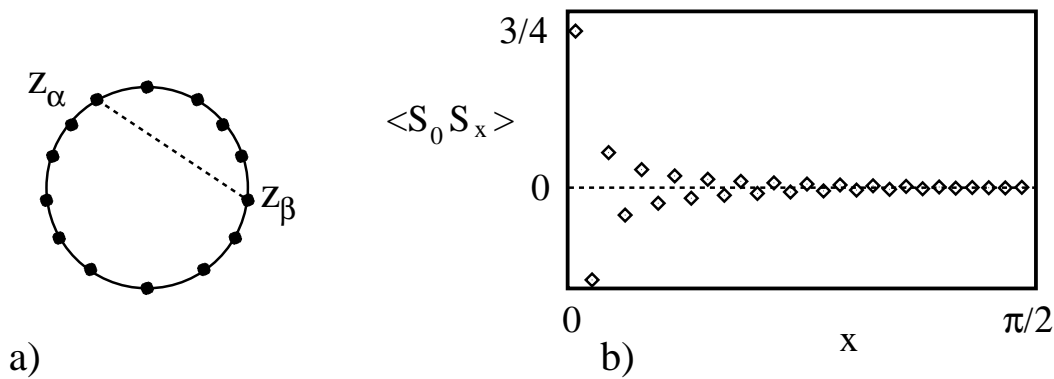}
% includes figure foo.eps
\caption{ {\bf a)} Haldane-Shastry interaction; {\bf b)} Spin-spin static
correlations in the ground state of the Haldane-Shastry model. }
\label{figu1}
\end{figure}

The HS model is a system of spin 1/2 particles lying at the sites of
a lattice with periodic boundary conditions. The lattice may be thought
of as ``wrapped'' around a circle. The interaction among the
spins is antiferromagnetic, with interaction strength $J$, and inversely
proportional to the square of the chord between the corresponding sites.
Let
$N$ be the number of lattice sites, so that a generic point on the lattice
will be labeled by $\alpha = 1 , \ldots , N$. Let us also define the
analytic coordinate on the lattice, $z_\alpha = \exp \biggl[  2 \pi i
\alpha /  N \biggr]$. The HS-model Hamiltonian reads:

\beq
{\cal H}_{HS} = \frac{J}{2} \left( \frac{ 2 \pi}{N} \right)^2
\sum_{\alpha \neq \beta = 1}^N \frac{ {\bf S}_\alpha \cdot {\bf S}_\beta}{
| z_\alpha - z_\beta |^2}
\label{eq3.1}
\eneq
\noindent
where ${\bf S}_\alpha$ is the spin at site $\alpha$ ($({\bf S}_\alpha)^2 =
3/4)$.

Due to the magic of analytic variables on the circle, $z_\alpha^* = 1/
 z_\alpha$, the interaction in Eq.(\ref{eq3.1}) is, in fact, an analytic
function of the $z_\alpha$'s. This allows us to work out the action of
${\cal H}_{HS}$ on many-body states, depending on the $z_\alpha$'s, in
terms of differential operators acting on the analytic extension of the
corresponding wavefunctions.

${\cal H}_{\rm HS} $ clearly has the global $SU (2)$-symmetry generated
by the total spin ${\bf S} = \sum_{\alpha=1}^N {\bf S}_\alpha$. It also
posses an additional symmetry, corresponding to total spin-current
conservation and generated by the operator ${\bf \Lambda}$, given by:

\beq
{\bf \Lambda} = \frac{i}{2} \sum_{\alpha < \beta = 1}^N \biggl( \frac{
z_\alpha + z_\beta}{ z_\alpha - z_\beta } \biggr) ( {\bf S}_\alpha
\times {\bf S}_\beta )
\label{eq3.2}
\eneq
\noindent
that is, one has:

\[
[ {\cal H}_{\rm HS} , {\bf S} ] = [ {\cal H}_{\rm HS} , {\bf \Lambda} ] =
0
\]
\noindent
The action of ${\cal H}_{\rm HS}$ on the ground state and on one- and
two-spinon collective states of the system will be the subject of the
remaining part of this Section.

\subsection{Ground state of ${\cal H}_{\rm HS}$}

Let $ |\Psi_{\rm GS} \rangle $ be the ground state of ${\cal H}_{\rm
HS}$.  Due to the low-dimensionality, in 1D quantum antiferromagnets
with short-range correlations, we cannot expect $ | \Psi_{\rm GS}
\rangle $ to order. Rather than being realized as a one-dimensional
antiferromagnetic state with Ne\'el order, the ground state of the
system is given by a one-dimensional spin-liquid state, with spin
density uniform over length scales of the order of the lattice
step. The ground state also has short-range correlations and no net
magnetization. For the HS model, in the even-$N$-case, $| \Psi_{\rm
GS} \rangle$ is expressed in terms of its components on states of the
Fock space with $N/2$ spin-$\uparrow$, the remaining ones being
$\downarrow$. In particular, let $ | z_1 , \ldots , z_{ N/2} \rangle $
be defined as:

\beq
| z_1 , \ldots , z_\frac{N}{2} \rangle = \prod_{ j = 1}^\frac{N}{2 }
S_j^+ | \downarrow , \ldots , \downarrow \rangle
\label{eq3.3}
\eneq
\noindent
where $ | \downarrow , \ldots , \downarrow \rangle$ is the fully polarized
states
with all the spins at the lattice sites being $\downarrow$, while $z_1 ,
\ldots , z_{N/2} $ are $N/2$-lattice coordinates. We define $ | \Psi_{\rm GS}
\rangle$ by:

\beq
\langle z_1 , \ldots , z_\frac{N}{2 } | \Psi_{\rm GS} \rangle =
\Psi_{\rm GS} ( z_1 , \ldots , z_\frac{N}{2} ) =
\prod_{i < j = 1}^\frac{N}{2} ( z_i - z_j )^2 \prod_{t = 1}^\frac{N}{2}
z_t
\label{eq3.4}
\eneq
\noindent
${\cal H}_{\rm HS}$ can be traded for a differential operator, when
acting on functions of the $z_\alpha$'s as discussed in detail, for
instance,
in \cite{chia}. Here we outline the main idea and briefly sketch the
basic steps of the derivation.

First of all, we have to split the scalar product ${\bf S}_\alpha \cdot
{\bf S}_\beta$ into $ ( S_\alpha^+ S_\beta^- +  S_\alpha^- S_\beta^+ ) / 2
+ S_\alpha^z S_\beta^z$ and to separately consider the action of the
various
operators on the ground state. Therefore, let us observe that
$[S_\alpha^+ S_\beta^- \Psi_{GS} ] (z_1 , \ldots , z_{M})$ is different
from
zero only if  one of the arguments $z_1 , \ldots , z_{N/2}$ equals
$z_\alpha$.
By Taylor expanding the corresponding function, we obtain:

\beq
[ \biggl\{ \sum_{\beta \neq \alpha}^{N} \frac{ S_{\alpha}^{+}
S_{\beta}^{-}}{\mid \! z_{\alpha} - z_{\beta} \! \mid^2 }
\biggr\} \Psi_{GS}] (z_{1}, \ldots , z_{M})
=  \sum_{\ell=0}^{ N -2} \sum_{j = 1}^M \frac{ z_j^{ \ell +1}}{\ell !} A_l
( \frac{ \partial^\ell}{ \partial z_j^\ell} ) \biggl\{
\frac{ \Psi_{GS} ( \{ z_j \} )}{  z_j} \biggr\} \;\;\; .
\label{eq3.5}
\eneq
\noindent
(where $\{z_j\}$ is a shorthand for $\{ z_1 , \ldots , z_M \}$).

The coefficients $A_l$ are calculated in \cite{chia,usprb1}. In
particular,
it is possible to show that they are all zero for  $N>l>2$. Therefore, we
can rewrite Eq.(\ref{eq3.5}) as:

\begin{displaymath}
 \sum_{j=1}^{M} \biggl\{ \frac{(N-1)(N-5)}{12} z_{j} - \frac{N-3}{2}
 z_{j}^2 \frac{\partial}{\partial z_{j}}
+ \frac{1}{2} z_{j}^3
 \frac{\partial^2} {\partial z_{j}^2} \biggr\}
  \biggl\{ \frac{ \Psi_{GS} (z_{1},
 \ldots , z_{M})}{ z_{j}} \biggr\}
\end{displaymath}

\beq
=\biggl\{ - \frac{N}{8} - \sum_{j \neq k}^{M}
\frac{1}{\mid \! z_{j} - z_{k} \! \mid^2 } \biggr\}
\Psi_{GS} (z_{1}, \ldots , z_{M}) \; \; \; .
\label{eq3.6}
\eneq
\noindent
where, in deriving Eq.(\ref{eq3.6}), we have used the identity:

\begin{displaymath}
\frac{z_{\alpha}^2}{(z_{\alpha}-z_{\beta})(z_{\alpha}-z_{\gamma})} +
\frac{z_{\beta}^2}{(z_{\beta}-z_{\alpha})(z_{\beta}-z_{\gamma})}
+ \frac{z_{\gamma}^2}{(z_{\gamma}-z_{\alpha})(z_{\gamma}-z_{\beta})} =
1 \;\;\; .
\end{displaymath}
\noindent
Finally, it is straightforward to prove that:

\beq
[ \biggl\{ \sum_{\beta \neq \alpha}^{N} \frac{ S_{\alpha}^{z}
S_{\beta}^{z}}{\mid \! z_{\alpha} - z_{\beta} \! \mid^2 }
\biggr\} \Psi_{GS}]  ( \{ z_j \} )
= \biggl\{ - \frac{N(N^2 - 1)}{48}
+ \sum_{j \neq k}^{M} \frac{1}{\mid \! z_{j} - z_{k} \!
\mid^2 } \biggr\}
\Psi_{GS} ( \{ z_j \} )
\; \; \; .
\label{eq3.7}
\eneq
\noindent
Eqs.(\ref{eq3.6},\ref{eq3.7}) lead us to the final result:

\beq
{\cal H}_{HS}  \Psi_{\rm GS} (z_1 , \ldots , z_\frac{N}{2} ) =
E_{\rm GS}  \Psi_{\rm GS} (z_1 , \ldots , z_\frac{N}{2} )
\label{eq3.8}
\eneq
\noindent
with:

\[
E_{\rm GS} = - J \biggl( \frac{\pi^2}{24} \biggr) \biggl( N + \frac{5}{N}
\biggr)
\]
\noindent
that shows that $\Psi_{GS} ( z_1 , \ldots , z_{M} )$ is, in fact, an
eigenstate of ${\cal H}_{\rm HS}$. By employing an approach similar to
the one leading to Eq.(\ref{eq3.8}), it is also possible to show that
$\Psi_{\rm GS}$ is a total spin singlet:$[ {\bf S} \Psi_{\rm GS} ] (
z_1 , \ldots , z_{M} ) = 0 $. The proof that $\Psi_{\rm GS} $ is the
only ground state of ${\cal H}_{\rm HS}$ is not immediate. It involves
the ``factorization property'' of ${\cal H}_{\rm HS}$, first
discovered by Shastry \cite{factoriz}, and discussed at length in
Ref.\cite{usprb1}. We will not prove it here, but will rather focus on
some relevant physical properties of the state described by $\Psi_{\rm
GS}$.

In Fig.\ref{figu1}{\bf b)}, we report the spin-spin correlations in
$\Psi_{\rm GS}$, $\langle \Psi_{\rm GS} | {\bf S}_\alpha \cdot {\bf
S}_\beta | \Psi_{\rm GS}\rangle$, as a function of the separation,
$\alpha -\beta$. We may clearly recognize the algebraic decay of spin
correlations which is expected for a spin-1/2 chain where the spectrum
of the elementary excitations on top of the ground state shows no gap
\cite{haldanegap}. As the total spin momentum of the state is zero
and the spin density is uniform we can refer to $\Psi_{\rm
GS}$ as to a disordered spin-singlet, homogeneous nondegenerate spin
liquid. Elementary excitations on top of such a spin-liquid
spin-singlet are provided by collective modes, carrying total spin
1/2, the spinons. The next subsection deals with these particles.

\subsection{One- and Two-spinon states of the Haldane-Shastry model}

Spinons are stable excitations of the spin-1/2 HS model. Therefore,
they maintain their integrity upon interaction. This means that, in a
scattering process, the numbers of incoming and outgoing spinons are
equal.  Therefore, it is possible to diagonalize ${\cal H}_{\rm HS}$
within a subspace at fixed spinon number. As the total spin
is also a constant of motion of the model, subspaces with fixed total
spin polarization of the spinons do not mix one with each other upon
interaction. This justifies the procedure we make use of in this
Section, diagonalizing ${\cal H}_{\rm HS}$ in the one-spinon
and in the two-spinon fully polarized subspaces, respectively.

In order to construct one-spinon states, let us consider the HS-chain with
an odd number of sites. In this case, the minimum possible spin for each
state
is be 1/2. In particular, the ground state of the system will now be a
superposition of states with total spin 1/2. The relevant spin-1/2 states
may be constructed from uniform spin-liquid states, by holding the spin
at a site $\alpha$ to be $\downarrow$, for instance. As a function of the
$\uparrow$ spin-coordinates, $z_1 , \ldots , z_{ M}$ ($M = ( N - 1 )/2$) ,
the  corresponding wavefunction reads:

\beq
\Psi_{\alpha}^{\rm sp}
 ( z_1 , \ldots , z_M ) = \prod_{ j = 1}^M ( z_j - z_\alpha )
\prod_{i < j = 1}^M ( z_i - z_j )^2 \prod_{t = 1}^M z_t
\label{eq3.9}
\eneq
\noindent

\begin{figure}
\includegraphics*[width=0.8\linewidth]{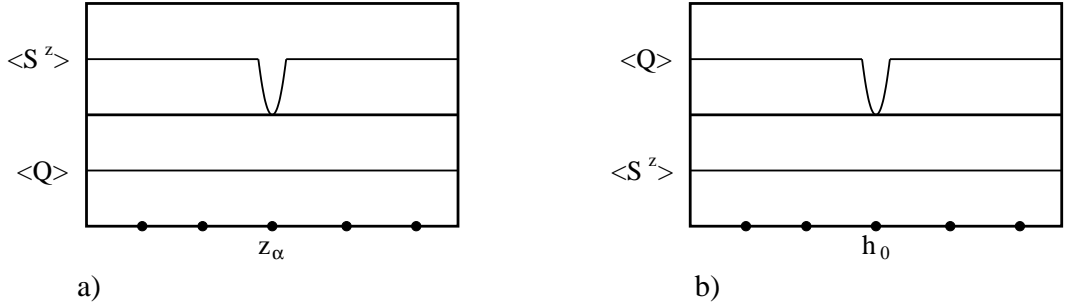}
% includes figure foo.eps
\caption{ {\bf a)} Profile of the spin- and charge-distribution of
a localized spinon at $z_\alpha$; {\bf b)} Profile of the spin- and
charge-distribution of a localized holon at $h_0$.}
\label{figu2}
\end{figure}

In Fig.\ref{figu2}{\bf a)}, we sketch the spin density and the charge
density
profile for the state in Eq.(\ref{eq3.9}). As it appears from the plot,
$\Psi_\alpha^{\rm sp}$ describes a
uniform spin-liquid state, just like $\Psi_{\rm GS}$ does, except for a
localized ``spin bump'' at $z_\alpha$, carrying total spin -1/2. Such a
spin
bump is  a ``localized $\downarrow$-spinon at $z_\alpha$''.

Localized one-spinon states are not translationally invariant. Therefore,
they cannot be eigenstates of the Haldane-Shastry Hamiltonian. Eigenstates
of ${\cal H}_{\rm HS}$ may be constructed, instead, by propagating the
spinon,
that is, by using functions in the form:

\beq
\Psi_m^{\rm sp} ( z_1 , \ldots , z_M ) = \frac{1}{N} \sum_{ \alpha = 1}^N
(z_\alpha^* )^m \Psi_\alpha^{\rm sp}  ( z_1 , \ldots , z_M )
\;\; ; \;
0 \leq m \leq \frac{N-1}{2}
\label{eq3.10}
\eneq
\noindent
Indeed, by following the same steps leading to Eq.(\ref{eq3.8}), it is
possible to show that the eigenvalue equation for $\Psi_\alpha^{\rm sp}$
becomes:

\[
{\cal H}_{\rm HS} \Psi_{\alpha}^{\rm sp} = E_{\rm GS}  \Psi_{\alpha}^{\rm sp}
+ \frac{J}{2} \left( \frac{ 2 \pi}{N} \right)^2
\biggl\{ M ( M - 1 ) - z_\alpha^2 \frac{ \partial^2}{ \partial z_\alpha^2}
\]

\beq
- \frac{N-3}{2} \biggl[ M - z_\alpha \frac{ \partial}{ \partial z_\alpha}
\biggr] \biggr\}  \Psi_{\alpha}^{\rm sp}  = \lambda  \Psi^{\rm
sp}_{\alpha}
\label{eq3.11}
\eneq
\noindent
>From Eqs.(\ref{eq3.10},\ref{eq3.11}), we derive the eigenvalue $E_m$,
corresponding to $\Psi_m$:

\[
E_m = - J \biggl( \frac{\pi^2}{24} \biggr) \biggl( N - \frac{7}{N} \biggr)
+ \frac{J}{2} \biggl( \frac{2 \pi}{N} \biggr)^2  m
( \frac{N-1}{2} - m )
\]

\beq
= E_{\rm GS} + \frac{J}{2} \biggl[ \biggl( \frac{\pi}{2} \biggr)^2 -
(q_m^{\rm sp})^2
\biggr]
\label{eq3.12}
\eneq
\noindent
where, in Eq.(\ref{eq3.12}), we have introduced the crystal momentum
$q_m^{\rm sp} = \pi N / 2 - 2 \pi ( m + 1/4 ) / N$.

\begin{figure}
\includegraphics*[width=0.8\linewidth]{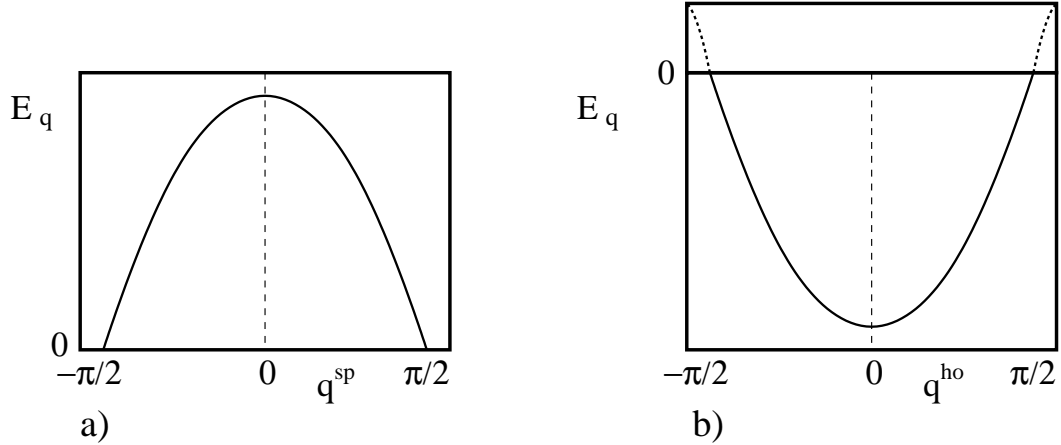}
% includes figure foo.eps
\caption{ {\bf a)} One-spinon dispersion relation;
{\bf b)} Negative-energy part of the one-holon dispersion relation.}
\label{figu3}
\end{figure}

Let us discuss some important features of the one-spinon dispersion
relation, $E ( q^{\rm sp}_m ) = \frac{J}{2} \biggl[ \biggl(
\frac{\pi}{2} \biggr)^2 - (q_m^{\rm sp})^2 \biggr]$. First of all, let
us notice that the one-spinon Brillouine zone (BZ) is halved, that is,
$-\pi / 2 \leq q_m^{\rm sp} \leq \pi / 2$.  This corresponds to the
absence of negative energy one-spinon states and,
correspondingly, to the halving of the number of available states.
Secondly, as we may see also from the plot of the dispersion relation
reported in Fig.\ref{figu3}{\bf a)}, the dispersion relation becomes
gapless at the endpoints of the BZ. The existence of gapless
excitations is a typical feature of half-odd spin antiferromagnets
\cite{haldanegap}, and is the reason for the algebraic decay of the
spin-spin
correlations we see in Fig.\ref{figu1}{\bf b)}.

Since spinons maintain their integrity, we can build multi-spinon states
by first creating $M$-localized spinons and then by propagating each one
of them. A residual spinon-interaction will require a further
diagonalization
of the Hamiltonian matrix restricted to the multi-spinon plane-wave
subspace,
as we will see in the particular case of two-spinon states.

To construct the state for two localized spinon at $z_\alpha$ and
$z_\beta$,
let us consider a HS-chain with even $N$. Let the $M ( = N/2 - 1)$ spins
at $z_1 , \ldots , z_M$ be $\uparrow$ and let the spin at $z_\alpha$ and
$z_\beta$ be held $\downarrow$. The corresponding collective wavefunction
will be given by:

\beq
\Psi_{\alpha \beta}^{\rm sp} ( z_1 , \ldots , z_M ) =
\prod_{ j = 1}^M (z_j - z_\alpha )
(z_j - z_\beta ) \prod_{ i < j =1}^M ( z_i - z_j )^2 \prod_{t = 1}^M z_t
\label{eq3.13}
\eneq
\noindent
A set of independent two-propagating-spinon wavefunctions is given by
the states:

\beq
\Psi_{m n}^{\rm sp}
  ( z_1 , \ldots , z_M ) = \frac{1}{ N^2} \sum_{ \alpha = 1}^N
\sum_{ \beta = 1}^N ( z_\alpha^* )^m ( z_\beta^* )^n
\Psi_{ \alpha \beta }^{\rm sp}
 ( z_1 , \ldots , z_M ) \;\; ; \; m \geq n = 0 \ldots M
\label{eq3.14}
\eneq
\noindent

The Schr\"odinger equation for $\Psi_{\alpha \beta}^{\rm sp}$ now reads:

\[
{\cal H}_{\rm HS}
\Psi^{\rm sp}_{\alpha \beta} = E_{\rm GS} \Psi^{\rm sp}_{\alpha \beta } +
\frac{J}{2} \biggl( \frac{ 2 \pi}{ N } \biggr)^2
\biggl\{ - \frac{z_\alpha^2}{
z_\alpha - z_\beta } \frac{ \partial}{ \partial z_\alpha } -
 \frac{z_\beta^2}{z_\beta - z_\alpha } \frac{ \partial}{ \partial z_\beta
}
- z_\alpha^2 \frac{ \partial^2}{ \partial z_\alpha^2} - z_\beta^2 \frac{
\partial^2}{ \partial z_\beta^2}
\]

\beq
 + \biggl( \frac{N-3}{2} \biggr)
\biggl[ z_\alpha \frac{ \partial}{ \partial z_\alpha} + z_\beta
\frac{ \partial}{ \partial z_\beta } \biggr] + [ 2 M^2 - M
( N - 2 ) ] \biggr\} \Psi^{\rm sp}_{ \alpha \beta } =
\lambda \Psi^{\rm sp}_{\alpha \beta }
\label{eq3.15}
\eneq
\noindent
In the plane-wave basis, Eq.(\ref{eq3.15}) becomes:

\[
\lambda \Psi_{mn}^{\rm sp} = \frac{J}{2} \biggl( \frac{2 \pi}{N} \biggr)^2
\biggl\{ - \frac{N^2}{48} \biggl( N - \frac{19}{N} + \frac{24}{N^2}
\biggr)
\]

\beq
+ m ( M - m ) + n ( M - n ) - \frac{m-n}{2} \biggr\} \Psi_{mn}^{\rm sp}
- \sum_{\ell = 1}^{\ell_M} (m-n+ 2 \ell ) \Psi^{\rm sp}_{m + \ell , n -
\ell}
\label{eq3.16}
\eneq
\noindent
where $\ell_M = n$ if $n+m<M$, $\ell_M =M-m$, if $n+m \geq M$.
Eq.(\ref{eq3.16}) has solutions of the form \cite{usletter1,usprb1}:

\beq
\Phi_{mn}^{\rm sp} =
\sum_{ \ell = 0}^{\ell_M} a_\ell^{mn} \Psi^{\rm sp}_{m + \ell, n - \ell}
\label{eq3.17}
\eneq
\noindent
where the coefficients $a_\ell^{mn}$ are recursively defined as:

\beq
a_\ell^{mn} = - \frac{ m - n + 2 \ell}{ 2 \ell ( \ell + m - n +
\frac{1}{2} )
} \sum_{ k = 0}^{\ell - 1 } a_k^{mn} \;\;\; a_0^{mn} = 1
\label{eq3.18}
\eneq
\noindent
The corresponding eigenvalues are given by:

\beq
E_{mn} = E_{\rm GS} + \frac{J}{2} \biggl( \frac{ 2 \pi}{N} \biggr)^2
\biggl[ m ( M - m ) + n ( M - n ) - \frac{ | m-n | }{2} \biggr]
\label{eq3.19}
\eneq
\noindent
Eq.(\ref{eq3.19}) shows that the energy of the two spinons is equal to the
ground state energy plus the energies of each single spinon plus an
interaction contribution that is subleading in the thermodynamic limit.
Indeed,
as $N \rightarrow \infty$, we obtain:

\beq
E_{mn} - E_{\rm GS} \rightarrow E ( q^{\rm sp}_m , q^{\rm sp}_n ) =
E ( q^{\rm sp}_m ) + E ( q^{\rm sp}_n )
\label{eq3.20}
\eneq
\noindent
The fact that, in the thermodynamic limit, the energy is additive does
not mean that the effects of spinon interaction disappear as $N \to
\infty$. Instead, as we are going to see in the next Section, spinon
interaction safely survives the thermodynamic limit, and appears to be the
driving mechanism of the low-energy physics of one-dimensional
antiferromagnets.

\section{Spinon dynamics in the Haldane-Shastry model}

In this Section we study spinon interaction and its properties. To do
so, we have to introduce a formalism that allows to treat spinons as
actual quantum mechanical particles.

In Section II, we referred to the states $\Phi^{\rm sp}_{mn}$ as the
two-spinon energy eigenstates, and to the states $\Psi^{\rm
sp}_{\alpha \beta }$ as the states for two localized spinons at
$z_\alpha$ and $z_\beta$. Therefore, it is always possible to
fully decompose any of the $\Psi^{\rm sp}_{\alpha \beta}$'s within the
set of the $\Phi^{\rm sp}_{mn}$'s:

\beq
\Psi_{\alpha \beta }^{\rm sp} = \sum_{ m = 0}^M \sum_{n = 0}^m z_\alpha^m
z_\beta^n {\cal P}_{mn}^{\rm sp} ( \frac{ z_\beta}{z_\alpha} )
\Phi^{\rm sp}_{mn}
\label{eq4.1}
\eneq
\noindent
where ${\cal P}_{mn}^{\rm sp} (z)$ is a polynomial in $z$ of degree $m-n$.

By definition, the coefficients of the combination in Eq.(\ref{eq4.1})
provide the coordinate representation for the wavefunction for two spinons
in a state of energy $E_{mn}$. The corresponding equation of motion for
${\cal P}_{mn}^{\rm sp}$ can be derived from the following identity chain
(which is a direct consequence of Eq.(\ref{eq3.15})):

\[
E_{mn} \langle \Phi^{\rm sp}_{mn} | \Psi^{\rm sp}_{\alpha \beta } \rangle
=
\langle \Phi^{\rm sp}_{mn} | {\cal H}_{\rm HS} | \Psi^{\rm sp}_{\alpha
\beta }
 \rangle
\]

\[
 = E_{\rm GS}  \langle \Phi^{\rm sp}_{mn} |
\Psi^{\rm sp}_{\alpha \beta } \rangle +  
 \rangle\frac{J}{2} \biggl( \frac{ 2 \pi}{N}
\biggr)^2 
 \biggl\{ ( M - z_\alpha
\frac{ \partial}{ \partial z_\alpha} ) z_\alpha \frac{ \partial}{ \partial
z_\alpha} 
\]

\beq
+ (M -z_\beta \frac{ \partial}{ \partial z_\beta } ) z_\beta
\frac{ \partial }{ \partial z_\beta} - \frac{1}{2} \biggl( \frac{ z_\alpha
+
z_\beta}{z_\alpha - z_\beta } \biggr) \biggl( z_\alpha \frac{ \partial}{
\partial z_\alpha} - z_\beta \frac{ \partial }{ \partial z_\beta } \biggr)
\biggr\}\langle \Phi^{\rm sp}_{mn} |
\Psi^{\rm sp}_{\alpha \beta }
\label{eq4.2}
\eneq
\noindent
In the differential operator at the r.h.s. of Eq.(\ref{eq4.2}) we
recognize the energy operators of the two spinons, plus an interaction
term that is velocity dependent and diverges as $1/x$ at short
separation $x$ between the two particles.

\begin{figure}
\includegraphics*[width=0.8\linewidth]{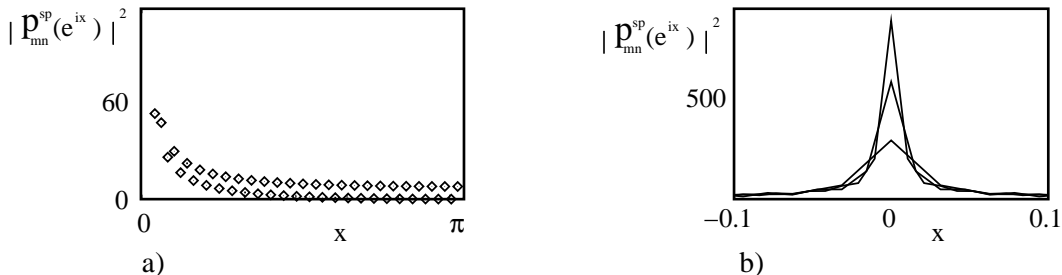}
% includes figure foo.eps
\caption{ {\bf a)} Square modulus of the two-spinon wavefunction for
$N=100$ and $m-n \sim N/2 -1$ as a function of spinon separation;
{\bf b)} Inset, at zero spinon separation, of the square modulus of the
two-spinon wavefunction for increasing $N$. From bottom to top:
$N=200,400,600$, respectively.}
\label{figu4}
\end{figure}

From the equation $\langle \Phi^{\rm sp}_{mn} |
\Psi^{\rm sp}_{\alpha \beta } \rangle$ =
$\langle \Phi^{\rm sp}_{mn} |
\Phi^{\rm sp}_{mn} \rangle z_\alpha^m z_\beta^n {\cal P}^{\rm sp}_{mn}
( \frac{z_\beta}{z_\alpha} )$ and from Eq.(\ref{eq4.2}), we may derive the
differential equation for the relative coordinate wavefunction:

\beq
\biggl\{ z ( 1 - z ) \frac{d^2}{d z^2} + \biggl[ \frac{1}{2} - m + n -
(\frac{3}{2} - m + n ) z  \biggr]  \frac{d}{d z } + \frac{m-n}{2}
\biggr\} {\cal P}_{mn}^{\rm sp} ( z ) = 0
\label{eq4.3}
\eneq
\noindent
Eq.(\ref{eq4.3}) is a degenerate hypergeometric equation, with parameters
$a = - m + n$, $b = 1/2$, $c = a + b$. Its solution is provided by the
hypergeometric polynomial \cite{abram}:

\beq
{\cal P}_{mn}^{\rm sp} ( z ) = {\cal P}_{m-n}^{\rm sp} ( z )
= \frac{ \Gamma [ m - n + 1] }{
\Gamma [ \frac{1}{2} ] \Gamma [ m - n +\frac{1}{2} ] } \sum_{k = 0}^{m-n}
\frac{ \Gamma [ k + \frac{1}{2} ] \Gamma [ m - n - k + \frac{1}{2} ]}{
\Gamma [ k + 1 ] \Gamma [ m - n - k + 1] } z^k
\label{eq4.4}
\eneq
\noindent
$ | {\cal P}^{\rm sp}_{mn} ( e^{i x} ) |^2$
gives the density of probability
for the configuration with two spinons at a distance $x$, as a function
of $x$. In Fig.\ref{figu4} we plot $ | {\cal P}_{mn}^{\rm sp} ( e^{i x} )
|^2$ vs. $x$. While at large separation the probability density
oscillates and averages to 1, as it is appropriate for noninteracting
particles, when the two spinons come close to each other, a huge
probability enhancement appears utill the function reaches its maximum for
$x=0$. The physical meaning of such a behavior may be summarized in two
basic points:

\begin{itemize}

\item {\bf Spinon interaction is short ranged}: This is clearly
seen when the probability amplitude approaches the one for two
noninteracting particles in the case when the spinons are far from
each other;

\item {\bf Spinon interaction is attractive}: This is seen in
the huge probability enhancement for configurations when the spinons
are on top of each other.

\end{itemize}

Therefore, we have identified spinon interaction as a short range
attraction. We now turn to the question of the fate of such an
interaction in the thermodynamic limit. In order to do so, let us look
at Fig.\ref{figu4}{\bf b)}, where we show an insert of the plot of $|
{\cal P}_{mn}^{\rm sp} ( e^{i x} ) |^2$ around $x=0$, for the same
relative momentum, but for increasing $N$. We see that, the larger
$N$ is, the stronger the enhancement. Therefore, not only spinon
interaction safely survives the thermodynamic limit but its effects
become more relevant as the the sample size grows.

\section{The Kuramoto-Yokoyama model of a one-dimensional insulator and
its
elementary excitations}

In this Section we discuss the Kuramoto-Yokoyama (KY) model for a
strongly correlated one-dimensional electron system. The KY-model is
the supersymmetric extension of the HS-model. Strong on-site
Coulomb repulsion forbids double occupancy, but one may have
fewer electrons than sites. This allows for charged excitations (holes).

The Kuramoto-Yokoyama Hamiltonian is defined as \cite{kuramoto}:

\[
{\cal H}_{\rm KY} = \frac{J}{2}  \biggl( \frac{ 2 \pi}{ N } \biggr)^2
\frac{1}{ | z_\alpha - z_\beta |^2} P_G \biggl\{ {\bf S}_\alpha \cdot
{\bf S}_\beta - \frac{1}{2} \sum_\sigma c_{\alpha \sigma}^\dagger c_{\beta
\sigma} + \frac{1}{2} ( n_\alpha + n_\beta ) - \frac{n_\alpha n_\beta}{4}
-\frac{3}{4} \biggr\} P_G
\]
\noindent
where $c_{\alpha \sigma}$ is the single electron annihilation operator,
 $n_\alpha = \sum_\sigma c_{\alpha \sigma}^\dagger
c_{\alpha \sigma} $ is the charge operator, and the Gutzwiller projector:

\[
P_G = \prod_{\alpha = 1}^N ( 1 - c_{\alpha \uparrow}^\dagger
c_{\alpha \downarrow }^\dagger c_{\alpha \downarrow} c_{\alpha \uparrow}
 )
\]
\noindent
accounts for the no-double occupancy constraint.

In addition to the usual ``bosonic'' symmetry, generated by the total spin
operator ${\bf S}$ and by the total charge operator $N = \sum_\alpha
n_\alpha$, ${\cal H}_{\rm KY}$ also possesses four
Fermionic symmetry operators,
given by ${\cal Q}_\sigma = \sum_{\alpha = 1}^N c_{\alpha \sigma}$ and by
${\cal Q}^\dagger_\sigma =  \sum_{\alpha = 1}^N c_{\alpha \sigma}^\dagger$
\cite{chia}.

 Exactly at half-filling, ${\cal H}_{\rm KY}$ reduces back to
${\cal H}_{\rm HS}$. Therefore, the ground state of ${\cal H}_{\rm KY}$ at
half-filling is still given by $ | \Psi_{\rm GS} \rangle$.
Also, charge-0, spin-1/2 spinon excitations will be described by the same
wavefunctions $\Psi_\alpha^{\rm sp}$ and $\Psi_m^{\rm sp}$ we have derived
for one spinon in the HS-model, with the same dispersion relation.

\subsection{One-holon and one-holon one-spinon states of the KY-model}

In order to construct charge-1, spin-0 holons, we may
consider a state where one electron is removed from the top of a spinon,
so that the resulting charge vacancy will carry no spin. The dispersion
relation of such an excitation spans, in general, the whole Brillouin zone.
Therefore, both positive- and negative-energy holons may arise, as charged
excitations of the KY-model. Because of the supersymmetry
of ${\cal H}_{\rm HS}$, a positive-energy holon state may be obtained by
acting with ${\cal Q}_\downarrow$ on one-spinon fully polarized
states. As those states will not be relevant for our analysis, we will
neglect them in the following, and will rather focus onto wavefunctions
with negative-energy holons.

In the case of odd $N$, the wavefunction for a propagating negative energy
holon is given by:

\beq
\Psi_n^{\rm ho} ( z_1 , \ldots , z_M | h ) = h^n \prod_{j = 1}^N (z_j - h)
\prod_{ i < j=1}^N ( z_i - z_j )^2 \prod_{ t = 1}^N z_t \;\; ; \;
0 \leq n \leq \frac{N+1}{2}
\label{eq5.2}
\eneq
\noindent
where $M= (N-1) / 2$, and $ z_1 , \ldots , z_M $ denote the position of
the
$\uparrow$-spin electrons. $h$ is the position of the empty site.

Clearly, $\Psi_n^{\rm ho}$ carries total charge 1. By using
the technique introduced for the HS-model, it is straightforward to prove
that $S^z \Psi_n^{\rm ho} = S^- \Psi_n^{\rm ho} = 0$. Therefore,
$S^z \Psi_n^{\rm ho}$ is a spin-0, charge-1 state, that is, a one-holon
spin singlet.

The mathematical derivation of the action of ${\cal H}_{\rm KY}$ on
 $\Psi_n^{\rm ho}$  is long and boring, although straightforward. Here, we
will not go through the various mathematical steps, but will rather quote
the final result:

\[
[ {\cal H}_{\rm KY} \Psi_n^{\rm ho}  ] ( z_1 , \ldots , z_M | h ) =
\biggl\{ - J \biggl( \frac{\pi^2}{24} \biggr) \biggl( N - \frac{1}{N}
\biggr) +
\frac{J}{2} \biggl( \frac{2 \pi}{N} \biggr)^2
\; n ( n - \frac{N+1}{2} ) \biggr\} \Psi_n^{\rm ho} ( z_1 , \ldots , z_M | h
)
\]

\beq
= \biggl\{ - J \biggl( \frac{\pi^2}{24} \biggr) \biggl( N + \frac{5}{N} -
\frac{3}{N^2} \biggr) + E (q_n^{\rm ho} ) ) \biggr\} \Psi_n^{\rm ho}
\label{eq5.3}
\eneq
\noindent
In Eq.(\ref{eq5.3}) we have introduced the one-holon crystal momentum,
$q_n^{\rm ho} = \pi N / 2 + 2 \pi ( n - 1 / 4 ) / N$ $({\rm mod} \; 2 \pi)$, 
and the dispersion relation:

\beq
 E (q_n^{\rm ho} ) = - \frac{J}{2} \biggl[ ( \frac{\pi}{2} )^2 -
(q_n^{\rm ho} )^2 \biggr]
\label{eq5.4}
\eneq
\noindent

 The negative-energy part of the dispersion relation in Eq.(\ref{eq5.4})
is plotted in Fig.\ref{figu3}({\bf b)}.
Again, one may recognize that the dispersion relation becomes gapless at
the
corners of the Brillouin zone, similarly to what happens for spinons. The
remaining half-part of the Brillouine zone is spanned by positive-energy
states, which we are not considering here.

The state for a localized holon at $h_0$, provided by the Fourier
transform of $\Psi_n^{\rm ho}$ reads:

\beq
\Psi_{h_0}^{\rm ho} = \sum_{n = 0}^\frac{N + 1}{2} h_0^{-n} \Psi_n^{\rm
ho}
\label{eq5.5}
\eneq
\noindent

Let us now introduce the one-spinon one-holon states of the KY-model.
The starting point is provided by the states $\Psi_{n,s}^{\rm sp,ho}$,
where the holon is propagating, while the spinon is localized at $s$.
Let $N$ be even, $M=N/2 - 1$, and let:

\[
\Psi_{n,s}^{\rm sp,ho} (z_1 , \ldots , z_M ) =
\]

\beq
h^n \prod_{j = 1}^M [ (z_j - s )
(z_j - h ) ] \prod_{ i > j = 1}^M (z_i - z_j )^2 \prod_{t = 1}^M z_t
\;\; ; \; 1 \leq n \leq M + 2
\label{eq5.6}
\eneq
\noindent
The KY-Hamiltonian acts on $\Psi_{n,s}^{\rm sp,ho}$ as follows
\cite{usletter2,usprb2}:

\[
[ \frac{ {\cal H}_{\rm KY}}{  \frac{J}{2} (\frac{2 \pi}{N} )^2 } - E_{\rm
GS} ]
\Psi_{n,s}^{\rm sp,ho}  (z_1 , \ldots , z_M ) =
\]

\[
\biggl\{ \biggl( n - M ) + ( M - 1 ) s \frac{ \partial}{ \partial s}
- s^2 \frac{ \partial^2}{ \partial s^2} + \frac{1}{2} \frac{ h + s}{ h -
s}
\biggl( s \frac{ \partial}{ \partial s} + ( 1 - n ) \biggr) \biggr\}
\Psi_{n,s}^{\rm sp,ho}
\]

\beq
+ \frac{ hs}{ ( h - s )^2} \biggl( \Psi_{n,s}^{\rm sp,ho} -
\left( \frac{s}{h} \right)^{n-1} \Psi_{n,s=h}^{\rm sp,ho}
\biggr)
\label{eq5.7}
\eneq
\noindent
In order to construct the corresponding energy eigenstates of
${\cal H}_{\rm KM}$, let us consider the one-spinon, one-holon plane
waves:

\beq
\Psi_{mn}^{\rm sp,ho}  (z_1 , \ldots , z_M )  = \sum_s \frac{ ( s^*)^m}{N}
\Psi_{n,s}^{\rm sp,ho}  (z_1 , \ldots , z_M )
\label{eq5.8}
\eneq
\noindent
In the plane wave basis, the eigenvalue equation takes the form:

\[
[ \frac{ {\cal H}_{\rm KY}}{ \frac{J}{2} ( \frac{2 \pi}{N} )^2 } - E_{\rm
GS} ]
\Psi_{mn}^{\rm sp,ho} = \biggl[ m ( M - m ) + n ( n - 1 + \frac{N}{2} )
\biggr] \Psi_{mn}^{\rm sp,ho}
\]

\beq
-\frac{1}{2} ( n - m + 1 ) \Psi_{mn}^{\rm sp,ho} - (n-m-1) \sum_{j=1}^m
\Psi_{m-j,n + j}^{\rm sp,ho}
\label{eq5.9}
\eneq
\noindent
if $m-n+1 < 0$, and:
\[
[ \frac{ {\cal H}_{\rm KY}}{ \frac{J}{2} ( \frac{2 \pi}{N} )^2 } - E_{\rm
GS} ]
\Psi_{mn}^{\rm sp,ho} = \biggl[ m ( M - m ) + n ( n - 1 + \frac{N}{2} )
\biggr] \Psi_{mn}^{\rm sp,ho}
\]

\beq
-\frac{1}{2} ( m - n + 1 ) \Psi_{mn}^{\rm sp,ho} - (m-n+1)
\sum_{j=1}^{M-m}
\Psi_{m+ j,n - j}^{\rm sp,ho}
\label{eq5.9bis}
\eneq
\noindent
if $m-n+1 \geq 0$.

Eqs.(\ref{eq5.9},\ref{eq5.9bis}) take the following eigenvectors and
eigenvalues:

\begin{itemize}

\item Case $m-n+1 < 0$:

\[
\Phi_{mn}^{\rm sp,ho} = \sum_{ j = 0}^m (-1)^j \frac{ \Gamma
[ j + \frac{1}{2} ]}{
\Gamma [ \frac{1}{2}] \Gamma [ j + 1 ] } \Psi^{\rm sp,ho}_{m - j , n + j}
\]

\beq
E_{mn}^+ = E_{\rm GS} + \frac{J}{2} \biggl( \frac{ 2 \pi}{N} \biggr)^2
\biggl[ m ( M - m ) + n ( n - 1 - \frac{N}{2} ) - \frac{1}{2} ( n - m - 1
)
\biggr]
\label{eq5.10}
\eneq
\noindent

\item Case $m-n+1 \geq 0$:

\[
\Phi^{\rm sp,ho}_{mn} = \sum_{ j = 0}^{M-m} (-1)^j \frac{
\Gamma [ j + \frac{1}{2} ]}{
\Gamma [ \frac{1}{2}]  \Gamma [ j + 1 ] } \Psi^{\rm sp,ho}_{m + j , n - j}
\]

\beq
E_{mn}^- = E_{\rm GS} + \frac{J}{2} \biggl( \frac{ 2 \pi}{N} \biggr)^2
\biggl[ m ( M - m ) + n ( n - 1 - \frac{N}{2} ) + \frac{1}{2} ( n - m - 1
)
\biggr]
\label{eq5.11}
\eneq
\noindent

\end{itemize}

Notice that, in either case, the energy can be written in terms of the
spinon and holon momenta as:

\beq
E_{mn} = E_{\rm GS} + E (q_m^{\rm sp} ) + E ( q_n^{\rm ho} ) -
\frac{\pi J}{N} | q_m^{\rm sp} - q_n^{\rm ho} |
\label{eq5.12}
\eneq
\noindent
 From Eq.(\ref{eq5.12}) we see that, as for the case of two spinons, in
the thermodynamic limit the energy of a spinon-holon pair on top of
the ground state energy reduces to the sum of the energies of the
single isolated spinon and of the single isolated holon. As for
spinons, this does not imply the interaction between spinons and
holons has no effects as the size of the system goes large.

\section{One-spinon one-holon dynamics in the Kuramoto-Yokoyama model}

In order to write down the Schr\"odinger equation for a spinon-holon
pair, Eq.(\ref{eq4.3}), in analogy with two spinons, we need to define
the state of a localized spinon at $s$ and a localized anyon at
$h_0$. Fourier transforming $\Psi_{n,s}^{\rm sp,ho}$ back to
coordinate space performs this task:

\beq
\Psi_{s , h_0}^{\rm sp, ho} ( z_1 , \ldots  , z_M ) = \sum_{n = 1}^{M+2}
h_0^{- n } \Psi_{n,s}^{\rm sp, ho}
\label{eq6.1}
\eneq
\noindent
$\Psi_{s, h_0}^{\rm sp, ho}$ can be fully decomposed within the basis of
one-spinon, one-holon energy eigenstates. In particular, we obtain:

\beq
\Psi_{s , h_0}^{\rm sp, ho} = \sum_{n=1}^{M+2} \sum_{m=0}^{n-2} s^m
h_0^{-n}
{\cal P}^{  \rm sp, ho}_{mn} ( \frac{s}{h_0} ) \Phi^{\rm sp, ho}_{mn}
+ \sum_{n=1}^{M + 2} \sum_{m = n-1}^{M}  s^m h_0^{-n}
{\cal P}^{  \rm sp, ho}_{mn} ( \frac{s}{h_0} ) \Phi^{\rm sp, ho}_{mn}
\label{eq6.2}
\eneq
\noindent
Eq.(\ref{eq6.2}) defines the one-spinon one-holon relative
wavefunction ${\cal P}_{mn}^{\rm sp, ho}$. To write down the
equation of motion for the one-spinon one-holon wavefunction, we need
the string of identities:

\[
( E_{mn}^{(\pm )} - E_{\rm GS} ) \langle \Phi_{mn}^{\rm sp, ho } |
\Psi_{s , h_0}^{\rm sp, ho} \rangle = \langle \Phi_{mn}^{\rm sp, ho} |
( \frac{{ \cal H}_{\rm KY}}{\frac{J}{2} ( \frac{ 2 \pi}{N} )^2 }
- E_{\rm GS} ) |  \Psi_{s , h_0}^{\rm sp, ho} \rangle =
\]

\[
\biggl\{ ( M - s \frac{ \partial}{ \partial s} ) s \frac{ \partial}{
\partial s} + ( \frac{N}{2} + 1 + h_0 \frac{ \partial}{\partial h_0} )
h_0 \frac{ \partial}{ \partial h_0}  + \frac{1}{2} \biggl( \frac{h_0 + s}{
h_0 - s } \biggr) \biggl( s \frac{ \partial}{ \partial s} + h_0 \frac{
\partial}{ \partial h_0 } + 1 \biggr) \biggr\}
 \langle \Phi_{mn}^{\rm sp, ho} |
\Psi_{s , h_0}^{\rm sp, ho} \rangle
\]

\beq
+ \frac{h_0}{s - h_0} \biggl( \frac{s}{h_0} \biggr)^\nu
 \langle \Phi_{mn}^{\rm sp, ho} |
\Psi_{s , h_0}^{\rm sp, ho} \rangle
\label{eq6.3}
\eneq
\noindent
where $\nu = M$ if $m-n+1 < 0$, $\nu = 0$ otherwise.

From Eq.(\ref{eq6.3}), it is straightforward to derive the differential
equation for ${\cal P}_{mn}^{\rm sp, ho}$:

\[
\biggl[ 2 \frac{d}{dz} - \frac{1}{(1-z)} \biggr] {\cal P}^{\rm sp, ho}_{
mn} ( z ) + \frac{ z^{M - m + 1}}{ (1-z)} {\cal P}^{\rm sp, ho}_{
mn} ( z ) = 0
\]
\noindent
if $m-n+1 < 0$, and:
\beq
\biggl[ 2 \frac{d}{d ( \frac{1}{z} ) } - \frac{1}{(1-\frac{1}{z} )}
\biggr]
{\cal P}^{\rm sp, ho}_{
mn} ( z ) + \frac{ ( \frac{1}{z})^{m}}{ (1-\frac{1}{z} )}
{\cal P}^{\rm sp, ho}_{
mn} ( z ) = 0
\label{eq6.4}
\eneq
\noindent
if $m-n+1 \geq 0$.

Notice that, unlike the spinon case, Eqs.(\ref{eq6.4})
are first-order differential equations. The reason for this is that
spinon and holon energy bands have opposite curvature and, because of
supersymmetry, the second derivatives of the dispersion relation have
equal
absolute values, but opposite signs. Nevertheless, one-spinon one-holon
dynamics is basically the same as two-spinon dynamics, as we are going to
discuss next.

\begin{figure}
\includegraphics*[width=0.8\linewidth]{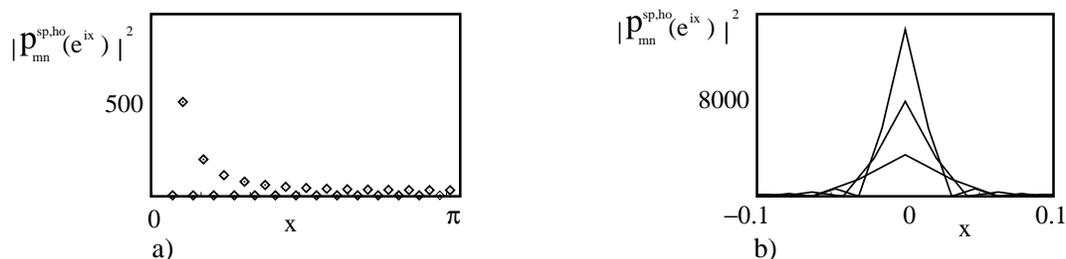}
% includes figure foo.eps
\caption{ {\bf a)} Square modulus of the one-spinon one-holon wavefunction
for
$N=100$ and $m-n \sim N/2 -1$ as a function of spinon-holon
separation; {\bf b)} Inset at zero spinon-holon separation of the
square modulus of the one-spinon one-holon wavefunction for increasing
$N$. From bottom to top: $N=100,200,400$, respectively.}
\label{figu5}
\end{figure}

Let us solve Eqs.(\ref{eq6.4}). It is easy to see that they allow for
polynomial-like solutions, given by:

\[
{\cal P}_{mn}^{\rm sp, ho} ( z ) = \sum_{k=0}^{M-m-1}
\frac{ \Gamma [ k + \frac{1}{2} ] }{ \Gamma [ \frac{1}{2} ] \Gamma [
k + 1 ] } z^k \;\;\; (m-n+1 < 0)
\]
and:

\beq
{\cal P}_{mn}^{\rm sp, ho} ( z ) = \sum_{k=0}^m
\frac{ \Gamma [ k + \frac{1}{2} ] }{ \Gamma [ \frac{1}{2} ] \Gamma [
k + 1 ] } ( \frac{1}{z} )^k \;\;\; (m-n+1 \geq 0).
\label{eq6.5}
\eneq
\noindent
The squared modulus of ${\cal P}^{\rm sp, ho}_{mn} ( e^{i x })$ gives
the probability of finding a spinon and a holon at a distance $x$ from
each other. Such a probability is plotted in Fig.\ref{figu5}{\bf
a)}. Notice that, as for a spinon pair, at large separations between
the two particles, the probability oscillates and averages to 1
flatly. This is a signal of the absence of long-range
interaction effects. As the two particles get close one to each other,
on the other hand, the probability peaks up, and shows a
consistent enhancement at zero separation.

Therefore an interaction between spinons and
holons exists. Such an interaction is short-ranged and, since it favors
configurations with the two particles on top of each other, it is
attractive. In conclusion, we assert that, as it happens among
spinons, there is a short-range attraction as a relevant interaction
between spinons and holons. The physical consequences of such an
attraction is the subject of the next Section.

\section{Physical consequences of the interaction between spinons and
holons}

In this Section we address the question of how spinon-spinon and
spinon-holon interaction can be seen in a real experiment.

The experiments we think about are neutron
scattering measurements  performed on quasi one-dimensional
antiferromagnets
\cite{alan}, and  ARPES spectra measurements, performed on quasi
one-dimensional insulators \cite{zxshen}. In the former case, one measures
the dynamical spin form factor, that is, the imaginary part of the spin
susceptibility, defined as:

\[
{\cal A}^{\rm spin}  ( \omega , q ) = \frac{1}{3} \Im m \{ \langle
\Psi_{\rm GS } | {\bf S} ( - \omega , - q ) \cdot {\bf S} ( \omega , q ) |
\Psi_{\rm GS}  \rangle \}
\]
\noindent
In the latter case, on the other hand, the photoemission of an electron
from
the sample is induced and, by detecting the emitted photoelectron,
one eventually reconstructs the spectral function of the recoiling
photohole,
defined as:

\[
{\cal A}^{\rm hole} ( \omega , q ) = \frac{1}{\pi}
\Im m \{ G_{\rm hole} ( \omega , q ) \}
\]
\noindent
where $G_{\rm hole} ( \omega , q )$ is the hole Green function.

By using the results we derived in the previous Section, we can
exactly calculate the contribution to ${\cal A}^{\rm spin} ( \omega ,
q )$ from two spinon states, as well as the contribution to ${\cal
A}^{\rm hole} ( \omega , q )$ coming from one-spinon one-holon states,
${\cal A}^{\rm sp, ho} ( \omega , q )$. In particular, we can express
these contributions in terms of the two-spinon and of the one-spinon
one-holon wavefunctions, calculated when the two particles are on top
of each other. In this way, we directly see what the measurable
consequences of the probability enhancement (and therefore the
interaction between particles) are.

The wavefunction for propagating  spin-one  excitation with momentum $q =
2 \pi k / N$ is given by:

\beq
S_q^- \Psi_{\rm GS} = \sum_\alpha (z_\alpha^*)^k 
\Psi_{\alpha \alpha}^{\rm sp} =
N \sum_{m=0}^N \sum_{n=0}^m (-1)^{m+n} z_\alpha^{m+n} \delta ( n + m - k )
{\cal P}^{\rm sp}_{mn} ( 1 ) \Phi_{mn}^{\rm sp}
\label{eq7.1}
\eneq
\noindent
Therefore, the form factor will be:

\[
{\cal A}^{\rm sp} ( \omega , q ) = 2 N^2 \Im m \biggl\{  \sum_{m=0}^N
\sum_{n=0}^m \frac{ \langle \Phi_{mn}^{\rm sp} | \Phi_{mn}^{\rm sp}
\rangle}{
\langle \Psi_{\rm GS} | \Psi_{\rm GS} \rangle } \times
\]

\beq
[ {\cal P}_{mn}^{\rm sp}
( 1 ) ]^2 \delta ( n + m - k ) \frac{ E_{mn} - E_{\rm GS}}{ ( \omega + i
0^+)^2
- ( E_{mn} - E_{\rm GS} )^2} \biggr\}
\label{eq7.2}
\eneq
\noindent
From Eq.(\ref{eq7.2}) we see that ${\cal A}^{\rm sp} ( \omega , q )$ is
fully
determined only by the squared modulus of the two-spinon wavefunctions at
zero separation  between the two spinons, that is, by the probability
enhancement. Since the probability enhancement is a direct consequence of
spinon attraction, we get to the ultimate result that the spin form
factor in one-dimensional antiferromagnets is fully determined only by
spinon attraction.

An expression similar to Eq.(\ref{eq7.2}) holds for ${\cal A}^{\rm sp, ho}
( \omega , q)$. In particular, we obtain \cite{usletter2,usprb2}:

\[
{\cal A}^{\rm sp,ho} ( \omega , q ) = \Im m \frac{1}{\pi} \biggl\{
\sum_{l=1}^{
M + 2} \sum_{m = 0}^{l - 2} \frac{ \delta ( k - m + l )
[ {\cal P}_{mn}^{\rm sp, ho} ( 1 ) ]^2 }{ \omega + i 0^+ - ( E^{+}_{mn}
- E_{\rm GS} ) }
\]

\beq
+  \sum_{l=1}^{
M + 2} \sum_{m = l-1}^{M} \frac{ \delta ( k - m + l )
[ {\cal P}_{mn}^{\rm sp, ho} ( 1 ) ]^2 }{ \omega + i 0^+ - ( E^{-}_{mn}
- E_{\rm GS} ) }
\label{eq7.4}
\eneq
\noindent
where, again, we see that the one-spinon one-holon contribution to the
hole spectral function is entirely determined only by the probability
enhancement, that is, by spinon-holon attraction.

The relevant quantities appearing in Eqs.(\ref{eq7.2},\ref{eq7.4}) are
the squared modulus of the wavefunctions when the two particles are on
top of each other, and the norms of the many-body two-spinon and
one-spinon one-holon states. The values of those quantities are reported
in Appendix A, where we show that they are basically given by products
and ratios of Euler's $\Gamma$-functions. Therefore, taking the
thermodynamic
limit of Eqs.(\ref{eq7.2},\ref{eq7.4}) is just matter of properly using
Stirling's approximation:

\[
\Gamma [ z ] \approx \sqrt{\pi} (z-1)^{z - \frac{1}{2}} e^{ - ( z - 1 )}
\;\;\; (z \; {\rm large } )
\]
\noindent
As discussed in detail in Refs.\cite{usprb1,usprb2}, the final result is:

\beq
{\cal A}^{\rm spin} ( \omega , q ) = \frac{J}{4}
\frac{ \Theta [ \omega_2 ( q ) - \omega ] \Theta [ \omega - \omega_{+1} (
q ) ]
\Theta [ \omega - \omega_{-1} ( q ) ] }{ \sqrt{ [ \omega - \omega_{-1} ( q
) ]}
\sqrt{ [ \omega - \omega_{ + 1 } ( q ) ] } }
\label{eq7.5}
\eneq
\noindent
(where $\omega_{-1} ( q ) = J/2 q ( \pi - q )$, $\omega_{+1} ( q ) =
J/2 ( 2 \pi - q ) ( q - \pi )$, and $\omega_2 ( q ) = 
J/2 q ( 2 \pi - q )$), for the spin form factor, and:

\beq
{\cal A}^{\rm sp,ho} ( \omega ,q ) = \frac{1}{ \pi^2 J q }
\sqrt{ \frac{ J [ q + \frac{\pi}{2} ]^2 - \omega}{ \omega - J [ q -
\frac{\pi}{2} ]^2 } }
\label{eq7.6}
\eneq
\noindent
for the one-spinon, one-holon contribution to the hole spectral density.

The quantities reported in Eqs.(\ref{eq7.5},\ref{eq7.6}) share some
important
features. Indeed,
we see that no resonances appear in either case. A resonance would be an
evidence for a stable spin-one propagating spin-wave in the former case,
and for a stable
propagating hole state in the latter case. We rather see broad features at
the tails of the spectra, basically signaling the nonintegrity of the
spin-wave
versus decay into a spinon pair, and of the hole versus decay into a
spinon-holon pair. Moreover, the spectra take sharp spikes at the creation
threshold, that is, when the two-spinon/spinon-holon pair is created at
the
minimum of energy. Such a sharp threshold appears quite a natural
consequence
of our results, that is, it is the effect of the attraction, taken in the
thermodynamic limit.

Sharp threshold features have been seen in the experiments quoted above
\cite{alan,zxshen}. They should be interpreted as a direct experimental
evidence of attraction among spinons and spinons and holons. Such an
attraction
appears to be ubiquitous for fractionalized excitations in one-dimensional
strongly correlated systems, despite the fact that we have derived it
within
the framework of a somehow oversimplified mathematical model.

\section{Conclusion}

In these notes we have reviewed some results concerning the
interaction among collective excitations of strongly correlated
one-dimensional electronic systems: holons and spinons. A detailed
mathematical derivation, performed within the framework of the
Haldane-Shastry and of the Kuramoto-Yokoyama model, the simplest
models of one-dimensional antiferromagnet and of one-dimensional
strongly correlated insulator, provided us with important informations
about such an interaction and its nature. We have seen that two
spinons, as well as a spinon and a holon, interact by means of a
short-range attraction. Although such an attraction creates a
probability enhancement that goes large in the thermodynamic limit, it
is not able to create a two-spinon or a one spinon-one holon bound
state (a propagating spin wave or a propagating hole,
respectively). The effect of the attraction is rather the creation of
a sharp emission threshold in the spin-wave and in the hole spectral
function, on top of a broad two-spinon, or one-spinon one-holon
continuum. Such a sharp threshold is seen in neutron scattering
experiment on quasi one-dimensional antiferromagnets \cite{alan}, and
in ARPES spectra measurements on quasi-one-dimensional insulator. In
our view, it provides relevant evidence for an attractive
interaction between spinons, and between spinons and holons.

\appendix

\section{Norms and probability enhancements}

In this Appendix we provide the formulas for the norms of two-spinon and
one-spinon, one-holon states and for the two-spinon and one-spinon,
one-holon
probability enhancements. We will skip the long and boring,
though straightforward, mathematical derivation of the results of this
Appendix, and refer the interested reader to the original papers
\cite{usprb1,usprb2}.

The starting point for the derivation of $\langle \Phi_{mn}^{\rm sp}
| \Phi_{mn}^{\rm sp} \rangle$ is a formula derived by K. Wilson
\cite{kwil}:

\beq
( \frac{ N}{2 \pi i } )^M \oint \frac{ d z_1}{z_1 } \ldots \oint \frac{ d
z_M}{
z_M} \prod_{i \neq j = 1}^M ( 1 - \frac{z_i}{z_j} )^2 =
N^M \frac{ ( 2 M )!}{ 2^M}
\label{appe1}
\eneq
\noindent
The next step is to consider the recursion relations

\[
\frac{ \langle \Phi_{mn}^{\rm sp} | \Phi_{mn}^{\rm sp} \rangle}{ \langle
 \Phi_{m-1, n}^{\rm sp} | \Phi_{m-1,n}^{\rm sp} \rangle}  =
\frac{ ( n - \frac{1}{2} ) ( M - n + \frac{3}{2} ) ( m - n + 1)^2}{
n ( M - n + 1 )( m - n + \frac{3}{2} )(m-n+\frac{1}{2} )}
\]
\noindent
and:

\beq
\frac{ \langle \Phi_{mn}^{\rm sp} | \Phi_{mn}^{\rm sp} \rangle}{ \langle
 \Phi_{m, n-1}^{\rm sp} | \Phi_{m,n-1}^{\rm sp} \rangle}  =
\frac{ ( M - m + 1 )( m-n+\frac{1}{2} )( m - n - \frac{1}{2} ) m}{
( M - m + \frac{1}{2} ) ( m + \frac{1}{2} ) ( m - n )^2}
\label{appe2}
\eneq
\noindent
Eqs.(\ref{appe2}) are solved by:

\[
\langle \Phi_{mn}^{\rm sp} | \Phi_{mn}^{\rm sp} \rangle =
C_M \frac{ \Gamma [ m - n + \frac{1}{2} ] \Gamma [ m - n + \frac{3}{2} ]
}{
\Gamma^2 [ m - n + 1 ] }
\frac{ \Gamma [ m + 1 ] \Gamma [ M - m + \frac{1}{2} ] }{ \Gamma [ m +
\frac{
3}{2} ] \Gamma [ M - m + 1 ] } \times
\]

\beq
\frac{ \Gamma [ n + \frac{1}{2} ] \Gamma [ M - n + 1 ]}{ \Gamma [ n + 1 ]
\Gamma [ M - n + \frac{3}{2} ] }
\label{appe3}
\eneq
\noindent
where $C_M$ may be calculated from Wilson's integral in Eq.(\ref{appe1}),
and
is given by $C_M = N^M (2 M )! (M+ \frac{1}{2} )/ ( \pi 2^M )$.

A similar technique may be applied, to derive $\langle \Phi_{mn}^{
\rm sp, ho} |\Phi_{mn}^{\rm sp, ho} \rangle$. The result is:

\[
\langle \Phi_{mn}^{\rm sp, ho} |\Phi_{mn}^{\rm sp, ho}
\rangle = N^{M+1} \frac{ ( 2 M )!}{ 2^M} ( M + \frac{1}{2} )
\frac{ \Gamma [ M - m + 1] \Gamma [ m + \frac{1}{2} ]}{ \Gamma [ M - m +
\frac{3}{2} ] \Gamma [ m + 1 ] }
\]
\noindent
for $m-n+1 <0$, and:

\beq
\langle \Phi_{mn}^{\rm sp, ho} |\Phi_{mn}^{\rm sp, ho}
\rangle = N^{M+1} \frac{ ( 2 M )!}{ 2^M} ( M + \frac{1}{2} )
\frac{ \Gamma [ M - m + \frac{1}{2} ] \Gamma [ m + 1 ]}{ \Gamma [ M - m +
1 ] \Gamma [ m + \frac{3}{2} ] }
\label{appe4}
\eneq
\noindent
for $m-n+1 \geq 0$.

The probability enhancement ${\cal P}_{mn}^{\rm sp} (1)$ may be derived
from standard properties of hypergeometric functions \cite{abram}. The
result is:

\beq
{\cal P}_{mn}^{\rm sp} (1) = \frac{ \Gamma [ \frac{1}{2} ] \Gamma [ m -n +
1]}{
\Gamma [ m - n + \frac{1}{2} ] }
\label{appe5}
\eneq
\noindent

Finally,  ${\cal P}_{mn}^{\rm sp,ho} (1)$ may be recursively derived, as
discussed in the Appendix of Ref.\cite{usprb2}, and the result is:

\[
{\cal P}_{mn}^{\rm sp,ho} (1) = 2 \frac{ \Gamma [ M - m +
\frac{1}{2}]}{ \Gamma [ \frac{1}{2} ] \Gamma [ M - m ] }
\]
\noindent
for $m-n+1 < 0$, and:

\beq
{\cal P}_{mn}^{  {\rm sp,ho}} (1) = 2 \frac{ \Gamma [ m +
\frac{3}{2} ]
}{ \Gamma [\frac{1}{2} ] \Gamma [ m  +1 ]}
\label{appe7}
\eneq
\noindent
for $m-n+1 \geq 0$.

\acknowledgments
This work is based on a lecture given by D. Giuliano at
the ``International School of Physics Enrico Fermi'' in Varenna, July
2002.
D. G. kindly acknowledges the organizers, A. Tagliacozzo, V. Tognetti and
B. Altschuler, for giving him the possibility to participate to the
stimulating
atmosphere of the school.

We ackowledge interesting discussions with A. Tagliacozzo and D. I.
Santiago.

\end{document}